\documentclass[preprint,showpacs,aps,prc,groupedaddress,floatfix]{revtex4}
\usepackage{epsfig}
\usepackage{amsmath,amssymb}
\usepackage{bm}

\newcommand{\hlf}{\mbox{$\frac{1}{2}$}}

\newcommand{\ri}{\mbox{${\rm i}$}}

\newcommand{\beq}{\begin{equation}}
\newcommand{\eeq}{\end{equation}}

\def\nuc#1#2{\relax\ifmmode{}^{#1}{\protect\text{#2}}\else${}^{#1}$#2\fi}

\begin{document}
\graphicspath{{figures/}}

\title{The  $l$ dependence of the helion optical model potential and the undularity of the $l$-independent equivalent}
\author{R. S. Mackintosh}
\email{r.mackintosh@open.ac.uk}
\affiliation{School of Physical Sciences, The Open University, Milton Keynes, MK7 6AA, UK}

\date{\today}

\begin{abstract}  The relationship between $l$-dependence  and undularity of the optical model potential (OMP)  is studied for the case of the elastic scattering of  33 MeV \nuc{3}{He}  on $^{58}$Ni. The relationship that emerges follows the general features of the same relationship found for the proton OMP presented in arXiv:1705.07003. The  general background is also presented in that reference.

\end{abstract}


\maketitle


\newpage
\section{INTRODUCTION}\label{intro}
Elsewhere, we have studied the relationship between the  $l$ dependence of the proton optical model potential (OMP) and the undularity (waviness) of the radial form of the $l$-independent  $S$-matrix equivalent, Ref.~\cite{rsm17}.  The $l$ dependence had been found to yield precise fits to cases of proton scattering for which precise and wide angular range data could not otherwise be fitted.  Here we pursue the same subject but for the case of  helion scattering. In what follows, an $l$-independent potential is said to be `$S$-matrix equivalent' to an explicitly $l$-dependent potential if it has an identical $S$-matrix $S_{lj}$. Such equivalent potentials are determined using 
$S_{lj} \rightarrow V(r) + {\bf l \cdot s}\, V_{\rm SO}(r) $ inversion, for which see Ref.~\cite{rsm17}.

\section{Evidence for $l$ dependence in \nuc{3}{He} scattering}\label{he3}
The angular distribution and analysing power for elastic scattering of \nuc{3}{He} at 33 MeV 
from \nuc{58}{Ni} could not be fitted by standard Woods-Saxon  phenomenology. In 
particular, the fit to the angular distribution was poor between $120^{\circ}$  and near $180^{\circ}$.
In Ref.~\cite{rm-he3}, the same data were analysed with the same form of $l$ 
dependence~\cite{cordero,kobmac79,kobmac81} that had been applied to proton scattering.

The $l$-dependent term, which was added to a standard 7-parameter WS plus WS-derivative 
$l$-independent central potential, had the following $l^2$-dependent form: 
\beq U_l(r) = f(l^2, L^2, \Delta^2) (V_l g_{\rm R}(r) + \ri W_l g_{\rm I} (r)) \label{lterm}\eeq
where the functions $g_{\rm R}(r)$ and $g_{\rm I}(r)$ are standard WS derivative
terms and $f(l^2, L^2, \Delta^2)$ is the standard WS, Fermi,  form with $L^2$ being the `radius' and $\Delta^2$
the `diffusivity'. The spin-orbit component had no $l$-dependent term. 

With this $l$-dependence, the 
 qualitative fit  to the angular distributions was greatly improved for $\theta \geq 120$ degrees
with $\chi^2/N$ halved. Subsequently, the same $l$ dependent model was applied,
Ref.~\cite{brum-he3}, to the scattering of polarised \nuc{3}{He} on \nuc{16}{O} and
\nuc{40}{Ca} but in this case the $l$-dependent component did not improve the fit.
The significant difference was that for both the \nuc{16}{O} and \nuc{40}{Ca}
cases the data terminated below 120 degrees. This is clear example of a case in which 
incompleteness of the data has concealed possible evidence for $l$ dependence. It is a 
pity since \nuc{16}{O} and \nuc{40}{Ca} are, like \nuc{58}{Ni},  of low collectivity, leading to 
elastic scattering angular distributions with well defined deep minima. Ref.~\cite{cage} presents
elastic scattering angular distributions (ADs) for 33 MeV \nuc{3}{He} on Ni isotopes of varying collectivity.
The backward angle fits for the least collective isotope, \nuc{58}{Ni}, with standard Woods-Saxon 
potentials, are much poorer than the fits for the more collective isotopes. We know this because of 
the relative completeness of the AD data for all these cases which extend to about $175^{\circ}$.  
This relationship between collectivity and ease of fitting applies to Ca isotopes for which  it was the least 
collective isotope, \nuc{40}{Ca}, that revealed the requirement for $l$ dependence in nucleon scattering. 
It appears that, for nuclei with low lying collective states, competing processes tend to wash out the 
sharp features in the ADs that make $l$-dependence manifest.

\section{Calculations \nuc{58}{Ni}}\label{58Ni}
After the introduction of $l$ dependence, the $S$-matrix  for high $l$ must not change, see Fig.~\ref{sm0} 
where it is clear that for $l$
even slightly above $L+  \Delta$ there is very little change in $S_{lj}$ in this figure and  Fig.~\ref{sm0-012}, the solid lines present the $S$-matrix with the $l$-dependence included, and the dashed lines without. The solid lines indicate an increase in absorption for the lowest partial waves and a decrease in $\arg S_{lj}$ (i.e. decrease in the real phase shift) corresponding to the small added repulsion for $l<L$.
Fig.~\ref{sm0-012} presents $S_{lj}$ for the lowest partial waves and the small reduction in $\arg S_{lj}$ due to the repulsive real $l$-dependent term can be made out more clearly.

\begin{figure}
\caption{\label{pot2-012} For 33 MeV \nuc{3}{He}  on $^{58}$Ni, the solid lines represent the $l$-independent part of the $l$-dependent potential and the dashed and dotted lines present inverted potentials, the dashed line with the larger inversion $\sigma$ corresponding to an earlier iteration. From the top the panels represent the real central, imaginary central, real spin-orbit and imaginary spin-orbit components. }
\begin{center}
\psfig{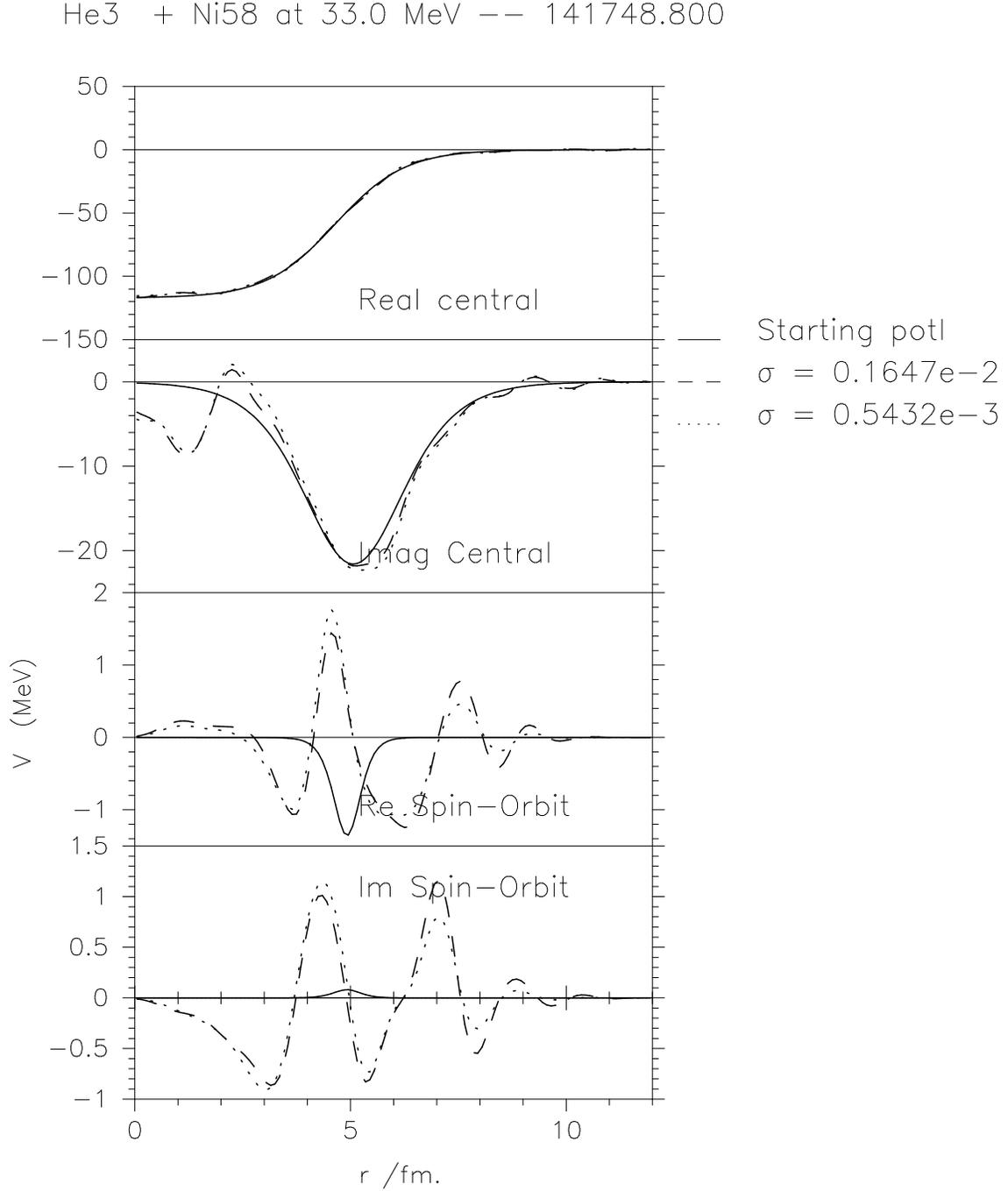}
\end{center}
\end{figure}

\begin{figure}
\caption{\label{pot4}  For 33 MeV \nuc{3}{He}  on $^{58}$Ni, the solid lines represent the $l$-independent part of the $l$-dependent potential and the dashed and dotted lines both represent the inverted potentials, after further iterations
and correspondingly lower $\sigma$ than those in Figure~\ref{pot2-012}.  From the top the panels represent the real central, imaginary central, real spin-orbit and imaginary spin-orbit components.}
\begin{center}
\psfig{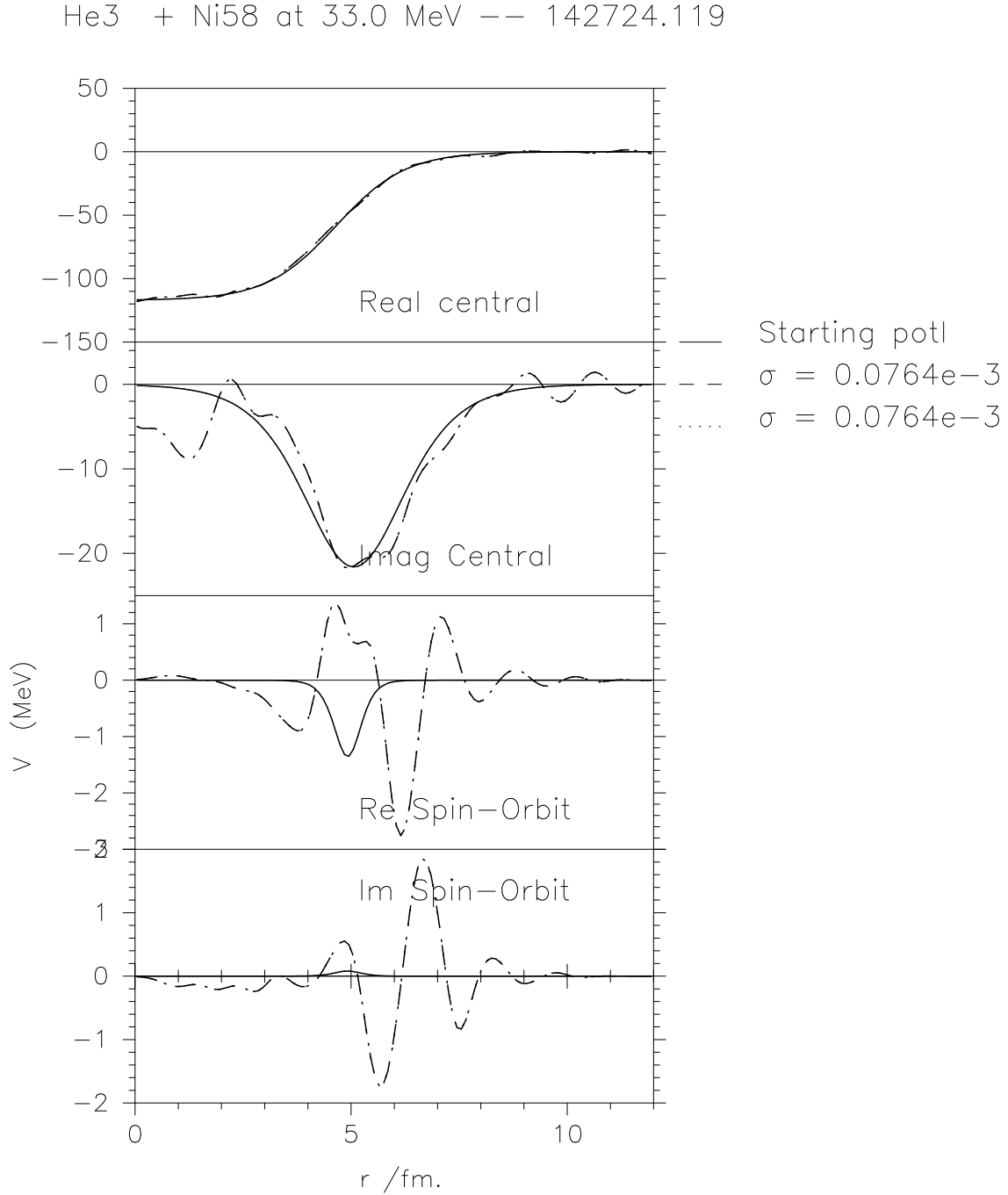}
\end{center}
\end{figure}

Fig.~\ref{pot2-012} presents two potentials that have been inverted by the IP method, showing how the degree of undularity increases as the inversion $\sigma$ falls. (The inversion $\sigma$ is defined in Ref.~\cite{rsm17}; the lower the value the more closely the inverted potential reproduces the $S_{lj}$.) In this figure, the solid lines present the $l$-independent potential to which the $l$-dependent terms have been added and the dashed and dotted lines present potentials that have been inverted from $S_{lj}$ calculated with the $l$-dependence included. The values of inversion $\sigma$ are,
$1.647 \times 10^{-3}$ and $ 5.432 \times 10^{-4}$ for the dashed and dotted curves. It is apparent that the real central part has not been changed much, and this is in accord with the small magnitude of the real $l$-dependent term in this case, and also in accord with the small effect on $\arg S_{lj}$  seen in Fig.~\ref{sm0-012}. However, a display of the real part on the same vertical scale as the imaginary part would reveal comparable undularities in the far surface. The shape of the imaginary central potential is essentially the same for the two solutions shown, and, in fact, the sharp bend for $r=0$ to $r= 4$ fm, is a well established property. That  includes the excursion into emissivity at 2 fm. It can be seen that the solution with the lower $\sigma$  (the dotted line) exhibits somewhat more undularity, a characteristic trend.
Pursuing further iterations of the inversion procedure improved the fit to $S_{lj}$ for high $l$, and the degree of undularity increases, see Fig.~\ref{pot4} which shows distinct emissivity at $r=9$ fm and $r=11$ fm with $\sigma = 7.64 \times 10^{-5}$.

Note that the regions of emissivity in the imaginary part of the inverted potential cannot result in a breach of the unitarity limit, $|S_{lj}| \le 1$ since the input $S_{lj}$ conform to that limit. 

\begin{figure}
\caption{\label{pot6a} For 33 MeV \nuc{3}{He}  on $^{58}$Ni, the solid lines represent the $l$-independent part of the $l$-dependent potential and the dashed and dotted lines present inverted potentials, the dashed line with the larger inversion $\sigma$ corresponding to an earlier iteration. In this case only the imaginary potential was $l$ dependent. From the top the panels represent the real central, imaginary central, real spin-orbit and imaginary spin-orbit components. }
\begin{center}
\psfig{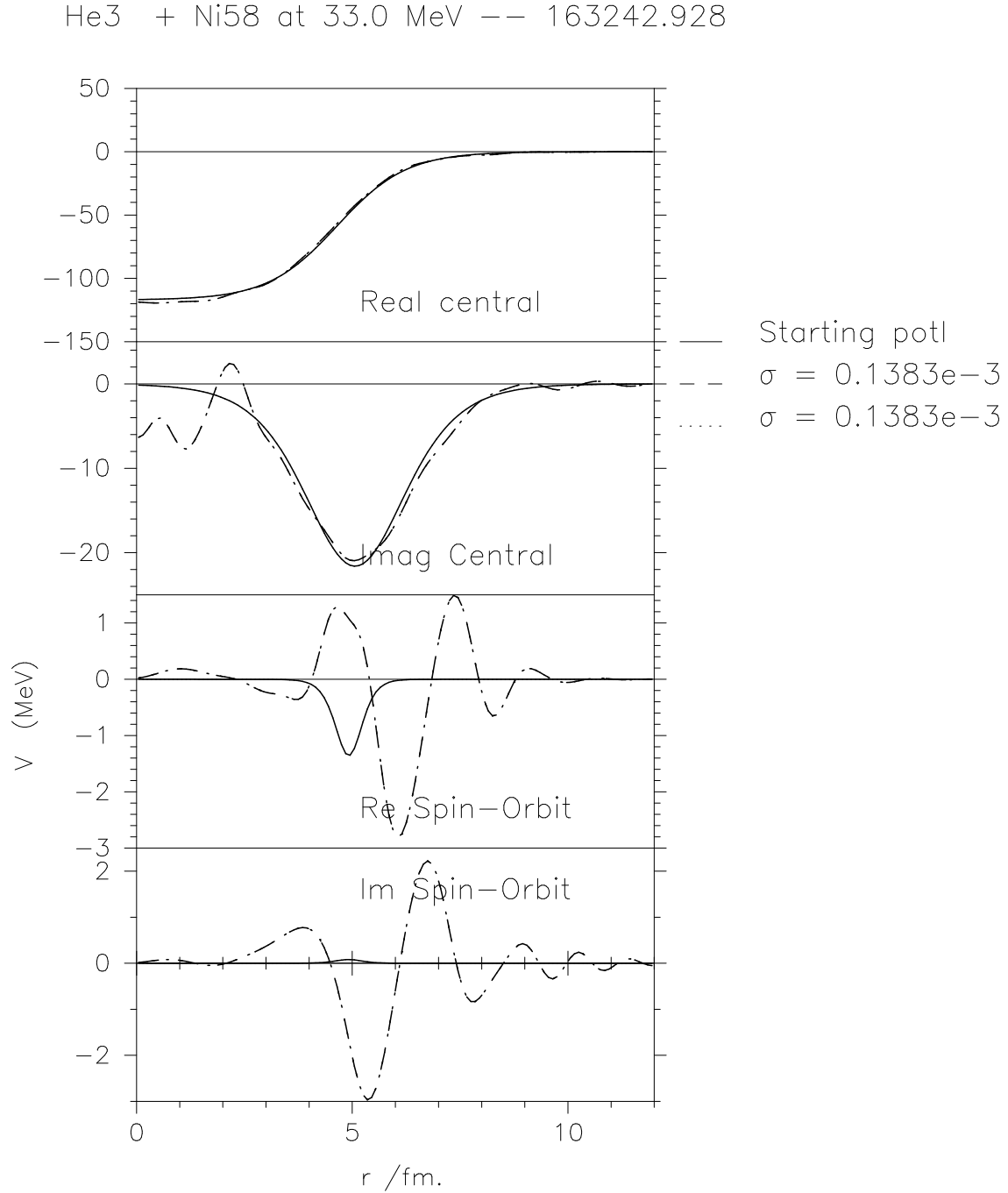}
\end{center}
\end{figure}

Fig.~\ref{pot6a} presents the inverted potential for the case with $l$ dependence for only the imaginary part of the potential. In line with the fact that the imaginary $l$ dependence  was stronger than the $l$ dependence in the real part, 
the main features found for the complete $l$-dependent potential re-appear, in particular the distinctive feature for $r<4$ fm in the imaginary part.

For both cases, the effects on the central potential are in line with what has been found in many cases involving $l$-dependence with proton scattering, but the strong effect on the spin-orbit terms was not expected. In this case, the original spin orbit term is small, with zero for the imaginary part, so the scale of the graph must be taken into account.  It is worth mentioning
that there is a long-recognized problem with the spin-orbit term for helion scattering, see for example Ref.~\cite{wjt}, one motivation studying the reaction channel contribution to the hellion OMP.

\section{Conclusions and discussion}\label{conc}
Evidently, the changes in the $l$-independent potential, that are required to reproduce the changes in $S_{lj}$ induced by the $l$ dependence, are of sufficient radial extent to influence the higher partial waves. 
It is therefore important to understand why the $l$ dependence, which can modify $S_{lj}$ only for partial waves having $l<L$ has such a conspicuous effect on the potential at large radii, well beyond the semi-classical radius corresponding to $L$.

Since the low-$l$ partial waves, that are influenced by the $l$-dependent term, will be sensitive to the potential at all radii, we must assume that the response to the $l$-dependent term  will be a modification of the potential at all radii. This would influence $S_{lj}$ for the high $l$  partial waves that are not directly affected by the $l$ dependent term, i.e.\ for which $l> L + \Delta$. The undularities then appear to be required to compensate for this and ensure that there is no change in $S_{lj}$ for all the higher partial waves.  Hence an $l$-dependent term that is active only for low partial waves must also induce, in the $l$-independent equivalent,  changes  in the potential at large radii, and that is what is found.

The natural conclusion is that the discovery of undulations in  OMPs, when experimental data are precisely fitted with model-independent methods, corresponds to a component in the potential that acts differently for high and low partial waves. Unfortunately, fitting of experimental data is often terminated at the appearance of undulations, perhaps because of the perceived lack of interpretations for such undularity.  Such undulations indicate an underlying $l$ dependency.

\begin{figure}
\caption{\label{sm0}  For 33 MeV \nuc{3}{He}  on $^{58}$Ni, the $S$-matrix  $S_{lj}$ with (solid lines) and without  (dashed lines) the $l$-dependent term. The upper panels are for $j = l - \hlf$ and the lower panels are for $j=l + \hlf$. In each pair, $|S_{lj}|$ is above and $\arg S_{lj}$ is below. The discontinuities in $\arg S_{lj}$
correspond to $2 \pi$ changes in the argument.}
\begin{center}
\psfig{figure=sm0.ps,width=15cm,angle=0,clip=}
\end{center}
\end{figure}

\begin{figure}
\caption{\label{sm0-012}  For 33 MeV \nuc{3}{He}  on $^{58}$Ni, the $S$-matrix  $S_{lj}$ with (solid lines) and without  (dashed lines) the $l$-dependent term, shown for $l=0$ to $l=12$. The upper panels are for $j = l - \hlf$ and the lower panels are for $j=l-+\hlf$. In each pair, $|S_{lj}|$ is above and $\arg S_{lj}$ is below. The discontinuities in $\arg S_{lj}$
correspond to $2 \pi$ changes in the argument.}
\begin{center}
\psfig{figure=sm0-012.ps,width=15cm,angle=0,clip=}
\end{center}
\end{figure}

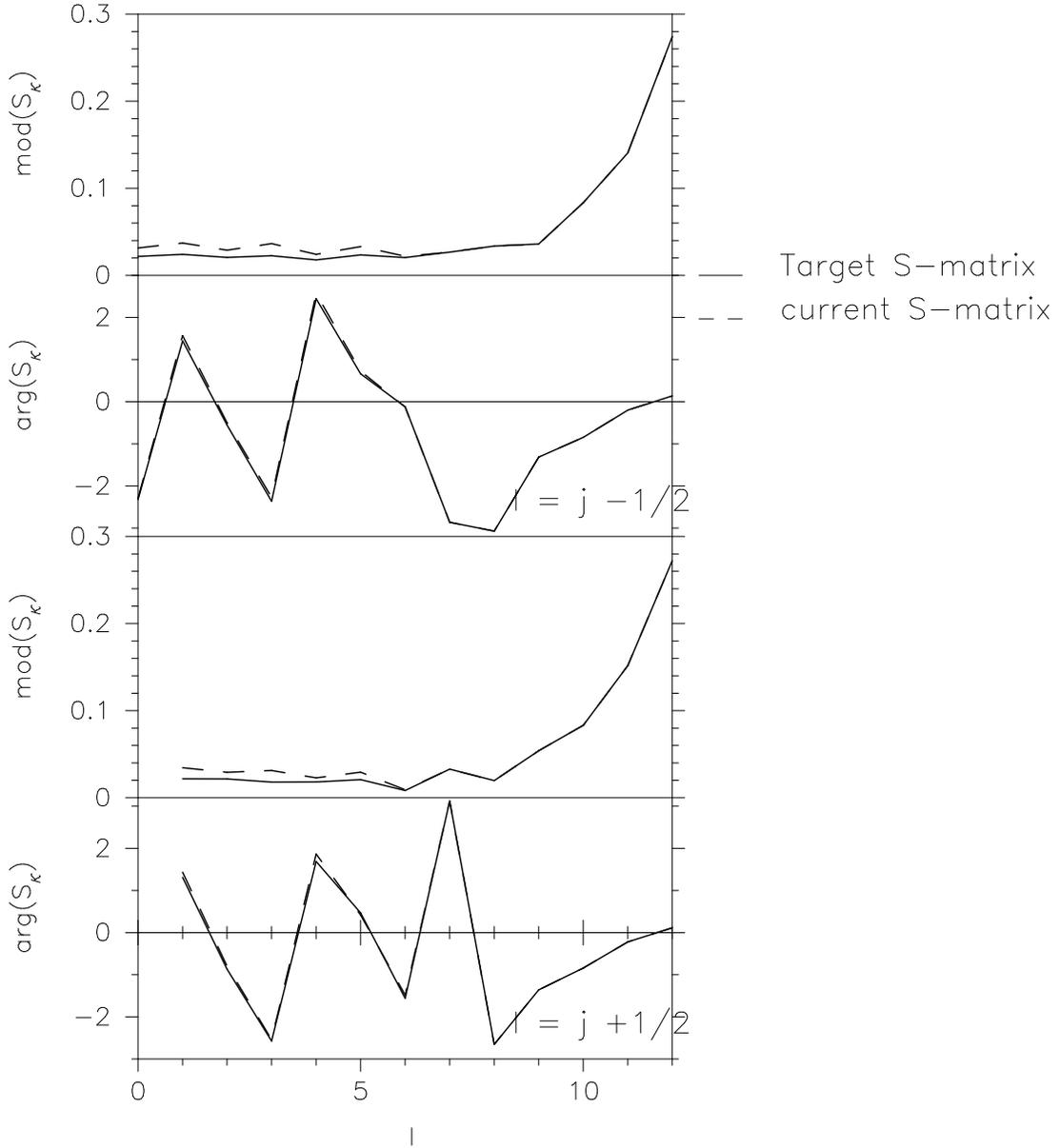
\end{document}